\documentclass[12pt]{article}
\usepackage{amsmath,amssymb,amsthm,cite}

\advance\topmargin by -\headheight
\advance\topmargin by -\headsep
\evensidemargin=\oddsidemargin
\renewcommand{\a}{\alpha}          %% abbreviation for \alpha
\renewcommand{\b}{\beta}           %% abbreviation for \beta
\newcommand{\C}{\mathbb{C}}        %% complex numbers
\newcommand{\cov}{{\mathrm{cov}}}  %% covariant
\newcommand{\Dl}{\Delta}           %% propagator
\newcommand{\del}{\partial}        %% abbreviation for \partial
\newcommand{\dl}{\delta}           %% abbreviation for Dirac's delta
\newcommand{\eps}{\epsilon}        %% abbreviation for \epsilon
\newcommand{\eucl}{{\mathrm{eucl}}}  %% Euclidean ('propagator')
\newcommand{\Ga}{\Gamma}           %% abbreviation for \Gamma
\newcommand{\K}{\mathcal{K}}       %% GLS functions
\newcommand{\La}{\Lambda}          %% abbreviation for \Lambda
\newcommand{\la}{\lambda}          %% abbreviation for \lambda
\newcommand{\loc}{{\mathrm{loc}}}  %% locally integrable
\newcommand{\nn}{\nonumber}        %% suppress equation numbering
\newcommand{\Om}{\Omega}           %% abbreviation for \Omega
\newcommand{\ox}{\otimes}          %% tensor product
\newcommand{\R}{\mathbb{R}}        %% real numbers
\DeclareMathOperator*{\Res}{Res}   %% residue
\newcommand{\ret}{{\mathrm{ret}}}  %% retarded (propagator)
\newcommand{\set}[1]{\{\,#1\,\}}   %% set notation
\newcommand{\cS}{\mathcal{S}}    %% Schwartz functions
\newcommand{\Sf}{\mathbb{S}}       %% S in `S-matrix'
\newcommand{\thalf}{\tfrac{1}{2}}  %% small fraction  1/2
\newcommand{\tquarter}{\tfrac{1}{4}} %% small fraction  1/4
\def\<#1,#2>{\langle#1,#2\rangle}  %% pairing of distributions

\makeatletter
\def\section{\@startsection{section}{1}{\z@}{-3.5ex plus -1ex minus
       -.2ex}{2.3ex plus .2ex}{\large\bf}}
\def\subsection{\@startsection{subsection}{2}{\z@}{-3.25ex plus -1ex
       minus -.2ex}{1.5ex plus .2ex}{\normalsize\bf}}
\makeatother

\theoremstyle{plain}
\theoremstyle{definition}
\begin{document}

\title{\bf Improved Epstein--Glaser renormalization\ \ II.\\
Lorentz invariant framework}

\author{Serge Lazzarini${\,}^1$\footnote{and also Universit\'e de la
M\'editerran\'ee, Aix-Marseille II,
e-mail\,: \texttt{sel@cpt.univ-mrs.fr}} \hskip 1.5mm and
Jos\'e M. Gracia-Bond\'{\i}a${\,}^2$\\[1cm]
${\,}^1$ Centre de Physique Th\'eorique\footnote{
Unit\'e Propre de Recherche 7061, et FRUMAM/F\'ed\'eration de Recherche 2291}
, CNRS--Luminy,\\
Case 907, F--13288 Marseille Cedex 9, France\\[5mm]
${\,}^2$ Departamento de F\'{\i}sica, Universidad de Costa Rica,\\
San Pedro 2060, Costa Rica
}
%%\date{}

\maketitle

\vskip 1cm

\hfill{To the Memory of Laurent Schwartz}

\vskip 1.5cm

% \newpage

\begin{abstract}
The Epstein--Glaser type $T$-subtraction introduced by one of the
authors in a previous paper is extended to the Lorentz invariant
framework. The advantage of using our subtraction instead of Epstein
and Glaser's standard $W$-subtraction method is especially important
when working in Minkowski space, as then the counterterms necessary to
keep Lorentz invariance are simplified. We show how
$T$-renormalization of primitive diagrams in the Lorentz invariant
framework directly relates to causal Riesz distributions. A covariant
subtraction rule in momentum space is found, sharply improving upon
the BPHZL method for massless theories.
\end{abstract}

\vskip 1cm

\noindent 2001 PACS Classification: 03.70.+k, 11.10.Gh, 11.30.Cp


\indent

\noindent{CPT--2002/P.4462}

\indent

\newpage

% \S 1
\section{Introduction}

According to the spirit of Epstein--Glaser theory~\cite{EG}, the physical
process of renormalization is mathematically expressed as an extension of
functionals, v.gr.\ (convolutions of) powers of Feynman propagators, to well
defined distributions. This paper is a continuation of~\cite{Carme} by the
second named author, in which an extension on configuration space was
presented. By modifying the use of, and relaxing the conditions on, the
infrared regulators $w$ introduced by Epstein and Glaser, very useful results
on the distributions at the crossroads of mathematics and quantum field theory
have been obtained; the relationship of this improved Epstein--Glaser
subtraction, dubbed ``$T$-renormalization'', with Hadamard regularization,
the minimal subtraction scheme in analytical regularization, and differential
renormalization, has been treated there at length. Hereinafter that
paper~\cite{Carme} will be denoted by~I.

The discussion in I took place in the Euclidean framework introduced
by Stora~\cite{EllRen} in the realm of the Epstein--Glaser
construction. We tackle in this paper the problem of going to the
``physical world'' with its symmetry group of transformations, namely
the Minkowski space and the Lorentz group respectively.

The Lorentz covariance properties of extensions for powers of
propagators are deeply related to the $S$-matrix covariance. Prima
facie, Epstein--Glaser procedures are not covariant. A proof of
existence of covariant time ordered products was first given
in~\cite{EG}, working on momentum space. About ten years later, the
problem was translated into a group cohomological question on
configuration space by Popineau and Stora in~\cite{PoSt} ---another
work which has remained at the status of preprint. The latter analysis
is available in textbook form~\cite{Scharf}. In relatively recent
(also apparently unpublished) preprints~\cite{Prangeetal, PrangeII},
explicit computations for ``counterterms'' reestablishing Lorentz
invariance of the extensions have been performed.

A perfectly covariant method for the extension of distributions employed in
quantum field theory exists, although it is rarely used: the ``analytical
regularization'' method~\cite{ThreeFromBA} of Bollini, Giambiagi and Gonzales
Dom\'{\i}nguez. It leads to quantum versions of the Riesz distributions of
classical field theory~\cite{Riesz, Jager, Kanwal}.

In this paper we extend $T$-renormalization of primitive diagrams to
the Lorentz-invariant framework, and show that it generalizes directly
the causal or quantum Riesz distributions.

The proof that all difficulties in renormalization theory can be overcome
through exclusive use of the $T$ operation is work in progress; the third
paper of the series~\cite{Flora} ---from now on denoted III--- deals with
renormalization of nonprimitive diagrams and the Hopf algebra of Feynman
graphs~\cite{DirkOriginal,CKI,Hektor}. It is of course not claimed that the
simplicity of the class of extensions envisaged in this work guarantees that
it is the physically relevant one in all circumstances; as a rule, the latter
will be singled out by appropriate renormalization prescriptions.

We do not suppose the reader to be familiar with causal Riesz distributions.
Thus we turn to them in Section~3, after recalling the main properties of our
``natural'' extension method in Section~2. The covariance properties of
$T$-renormalization are then addressed in Section~4. To fix ideas, we mainly
consider in Section~5 the basic primitively divergent graphs in massless
$\varphi^4_4$ theory, and demonstrate by way of example the link to analytical
regularization and causal Riesz distributions. Section 6 shows how
$T$-renormalization improves upon the BPHZL formalism. We conclude with a
brief discussion and outlook.

% \S 2
\section{Main features of $T$-renormalization}

Let us begin by fixing some conventions. The scalar functionals we
shall be dealing with (coming from the Wick theorem) are to be
extended to the main diagonal of~$\R^{4(n+1)}$. The latter can be
regarded as the origin in $\R^{4n}$ thanks to translation invariance,
allowing to set $x_{n+1} = 0$, for instance.

For fixed $j\in\{1,\dots,n\}$, to each fourcoordinate $x_j$ (a point
in $\R^4$) let $\a^j=(\a^j_0,\a^j_1,\a^j_2,\a^j_3)$ be a quadri-index,
where $\a^j_\mu$ is a nonnegative integer for each $j$ and each
Lorentz index $\mu$. One has, according to Schwartz's
notation~\cite{Schw}, $\a^j! = \a^j_0!\a^j_1!\a^j_2!\a^j_3!$ and
$|\a^j| = \a^j_0 + \a^j_1 + \a^j_2 + \a^j_3$, as well as
$$
x_j^{\a^j} := \prod_{\mu=0}^3 (x^\mu_j)^{\a^j_\mu},  \qquad
\del_{\a^j} := \left(\frac{\del}{\del x_j}\right)^{\a^j} =
\prod_{\mu=0}^3 \left(\frac{\del}{\del x^\mu_j}\right)^{\a^j_\mu} =:
\prod_{\mu=0}^3 \left(\del^j_\mu\right)^{\a^j_\mu}.
$$
We also use a multiquadri-index $\a=(\a^j_\mu)$ notation:
$\a! = \a^1!\a^2!\dots\a^n!$ and
$|\a| = |\a^1| + |\a^2| +\cdots+ |\a^n|$, as well as
$$
x^\a    := \prod_{j=1}^n x_j^{\a^j}, \qquad
\del_\a := \prod_{j=1}^n \del_{\a^j}.
$$
We assume that the reader is familiar with the basic concepts of
distribution theory. In this paper operators (Fourier transforms,
actions of the Lorentz group, subtractions, \dots) are defined on
distributions always by transposition (or adjoint mapping), and we
denote them by the \textit{same} letter denoting the original action
on test functions.

\smallskip

Consider scalar functionals $f(x_j)$ defined on
$M_4^{n+1} \setminus D_{n+1}$, where
$D_{n+1} := \{x_1 =\cdots= x_{n+1}\}$ is the full diagonal; thus we
assume that the diagram is primitive or that Bogoliubov's~\cite{IZ}
disentangling operation $\bar R$ corresponding to all subgraphs
(see~III) has been performed. Those will be considered as functionals
of the difference variables in $\R^{4n} \setminus \{0\}$. A tempered
distribution $\tilde{f}\in\cS'(\R^{4n})$ is an \textit{extension} or
\textit{renormalization} of~$f$ if
$$
\tilde{f}[\phi] \equiv \<\tilde{f},\phi> =
\int_{\R^{4n}} f(x)\phi(x) \,d^{4n}x
$$
holds whenever $\phi$ belongs to $\cS(\R^{4n}\setminus\{0\})$. In QFT
one considers a generalized homogeneity degree, the \textit{scaling
degree}~\cite{OS}. The scaling degree $\sigma$ of a scalar
distribution $f$ at the origin of $\R^d$ is defined to be
$$
\sigma(f) = \inf \set{s : \lim_{\la\to 0} \la^s f(\la x) = 0}
\qquad\mbox{for } f\in \cS'(\R^{4n}),
$$
where the limit is taken in the sense of distributions. Essentially
this means that $f(x) = O(|x|^{-\sigma(f)})$ as $x \to 0$ in the
Ces\`aro or distributional average sense~\cite{Ricardo}.

Let then $\sigma(f) = a$, with $a$ an integer, and $k = a - 4n\geq 0$.
Then, $f \notin L_\loc^1(\R^{4n})$. The simplest way to get an
extension of $f$ would appear to be standard Taylor series surgery:
throw away the $k$-jet $j^k_0\phi$ of $\phi$ at the origin, and define
$\tilde f$ by transposition:
$$
\<\tilde{f},\phi> = \<R^{k}_0f,\phi> := \<f,R^{k}_0\phi>,
$$
where $R^k_0\phi := \phi - j^k_0\phi$ is the Taylor remainder. Using
Lagrange's integral formula for $R^k_0$, and exchanging
integrations, one appears to obtain an explicit integral formula for
$R^k_0f$:
\begin{equation}
R^k_0f(x) \equiv T_1f(x) := (-)^{k+1} (k+1) \sum_{|\b|=k+1}
\del_\b \biggl[
\frac{x^\b}{\b!} \int_0^1 dt\,\frac{(1-t)^k}{t^{k+4n+1}}
f\Bigl(\frac{x}{t}\Bigr) \biggr].
\label{eq:bas-Lagr}
\end{equation}

The trouble with~\eqref{eq:bas-Lagr} is that the remainder $R^k_0\phi$
is not a test function and therefore, unless the infrared behaviour of
$f$ is good, we can end up with an undefined integral. Actually, for
the needs of theories with only massive fields,
formula~\eqref{eq:bas-Lagr} is largely sufficient. However in theories
with massless particles, $f$ is typically an homogeneous function with
an algebraic singularity, the infrared behaviour is pretty bad, and
$-4n$ is also the critical degree. A way to avoid the problem is to
\textit{weight} the Taylor subtraction. Epstein and Glaser~\cite{EG}
introduced infrared regulators $w$ with the properties $w(0) = 1$ and
$w^{(\a)}(0) = 0$ for $0 < |\a| \leq k$, as well as projector maps
$\phi\mapsto W_w\phi$ on $\cS(\R^{d})$ given by
\begin{equation}
W_w\phi(x) := \phi(x) - w(x) j^k_0\phi (x).
\label{eq:bas-EG}
\end{equation}

\smallskip

There is a considerable amount of overkill in~\eqref{eq:bas-EG}. We
argued in~I that one can, and should, weight only the \textit{last}
term of the Taylor expansion. This leads to the definition used in
this paper, at variance with Epstein and Glaser's:
\begin{equation}
T_w\phi(x) :=  \phi(x) - j^{k-1}_0(\phi)(x) - w(x) \sum_{|\a|= k}
\frac{x^{\a}}{\a!} \, \phi^{(\a)} (0).
\label{eq:imp-EG}
\end{equation}
Just $w(0) = 1$ is now required in principle for the weight function.
$T_w$ is also a projector. To obtain an integral formula for it, start
from
$$
T_w\phi = (1-w) R^{k-1}_0\phi + w R^k_0\phi.
$$
By transposition, using~\eqref{eq:bas-Lagr}, we derive
\begin{align}
T_w f(x)
&= (-)^{k} k \sum_{|\a|=k} \del_\a \biggl[ \frac{x^\a}{\a!}
\int_0^1 dt\, \frac{(1-t)^{k-1}}{t^{k+d}} f\Bigl(\frac{x}{t}\Bigr)
\Bigl(1-w\Bigl(\frac{x}{t}\Bigr)\Bigr)\biggr]
\nn \\
&\qquad + (-)^{k+1} (k+1)
\sum_{|\b|=k+1} \del_\b \biggl[ \frac{x^\b}{\b!}
\int_0^1 dt\, \frac{(1-t)^k}{t^{k+d+1}} f\Bigl(\frac{x}{t}\Bigr)
w\Bigl(\frac{x}{t}\Bigr) \biggr].
\label{eq:tapadelperol}
\end{align}

Consider the functional variation of the renormalized amplitudes with
respect to~$w$. One has
$$
\Bigl< \frac{\dl}{\dl w}T_w f, \psi \Bigr> :=
\frac{d}{d\la} T_{w+\la\psi}\,f\,\Bigr|_{\la=0}.
$$
Equation~\eqref{eq:imp-EG} yields:
\begin{equation}
\frac{\dl}{\dl w}T_w f[\cdot] = (-)^{k+1}
\sum_{|\a| = k} f[x^\a\cdot]\, \frac{\del_\a\dl}{\a!}\,,
\label{eq:funct-three}
\end{equation}
independently of $w$. The combination
$\frac{(-)^{|\a|}}{\a!}\,\del_\a\dl$ is rebaptized $\dl_\a$. A central
fact of renormalization theory, under its distributional guise, is
that there is no unique way to construct the renormalized amplitudes,
the inherent ambiguity being represented by the undetermined
coefficients of the $\delta$ and its derivatives, describing how the
chosen extension acts on the (finite codimension) space of test
functions which do \textit{not} vanish to some order in a neighborhood
of~$0$. There is, however, a more natural way ---in which the
ambiguity is reduced to terms in the higher-order derivatives of
$\delta$, as seen in~\eqref{eq:funct-three}, exclusively. This is
guaranteed by our choice of~$T_w$.

A word on the space of infrared regulators $w$ is in order. In I it
was shown that, for the extension of homogeneous $f$ of the kind found
in massless field models, any element of the space $\K'$ (dual of the
Grossmann--Loupias--Stein function space) of distributions rapidly
decreasing in the Ces\`aro sense~\cite{Odysseus}, taking the value~1
at zero, qualifies as a weight ``function''. The space $\K'$ is a kind
of distributional analogue of the Schwartz space $\cS$. Elements of
$\K'$ have moments of all orders. In particular, exponential functions
$e^{iqx}$ do qualify; this was realized by Prange, at the heuristical
level~\cite{PrangeA}. See the consequences in Section~6. The
usefulness of $\K'$ has appeared by now in many different
contexts~\cite{Hungarian}. The regulator $w(\mu x) = H(1-\mu|x|)$
of~I, with $H$ the Heaviside function, will be mainly used here. Call
$T_\mu$ the corresponding renormalization.

The results just summarized go a long way to justify the conjecture
(made by Connes and independently by Estrada) that Hadamard's finite
part theory is in principle enough to deal with quantum field theory
divergences. At least in the Euclidean context. When going from there
to the physical signature for the spacetime, both the ultraviolet and
the infrared problems immediately turn nastier, in a tangled sort of
way. For the first, since the singular support of the Feynman
propagator lies on the whole light-cone, it would appear that we have
to worry about singularities supported on the entire cone, and not
just at the origin. For the second, it is easy to see that if we
approach infinity in directions parallel to the light cone, the
Feynman propagator decays as $1/|x|$ and not ``na\"{\i}vely'' anymore
as~$1/|x|^2$.

Both kinds of trouble are a bit less ferocious than they seem. There
are techniques for dealing with the worsened infrared problem,
conjuring at need combinations of diagrams~\cite{OSphi}. Microlocal
analysis~\cite[Sec.~8.2]{Hoermander} can be invoked~\cite{Gudrun} to
argue that the ultraviolet troubles remain concentrated at the origin,
just as in the Euclidean case. In the next section, we show the same
for the square and the cube of the Feynman propagator (i.e., the basic
four-point and two-point divergent graphs in the $\varphi^4_4$ model),
by a direct calculation.

Whereas $T_1$ of~\eqref{eq:bas-Lagr} ostensibly preserves the Lorentz
covariance properties of $f$, the operator $T_w$
of~\eqref{eq:tapadelperol} in general does not. Our main task is to
fix this problem.

% \S 3
\section{Causal Riesz distributions}

Riesz's method consists in generalizing to the Lorentz-invariant context the
well known holomorphic family of distributions on $\R$,
$$
\Phi^\la(x) := \frac{x_+^{\la-1}}{\Ga(\la)}
$$
for complex $\la$, which have the properties
$$
\Phi^\la * \Phi^\mu  = \Phi^{\la+\mu}; \quad
\frac{d\Phi^\la}{dx} = \Phi^{\la-1}; \quad
\Phi^0(x) = \dl(x),
$$
where $*$ denotes convolution of distributions. In fact, Riesz only
dealt with the (advanced and) retarded propagators. He was able to
show that the holomorphic family of distributions $G^\la_\ret$ defined
on $\R^4$ as follows:
$$
G^\la_\ret(x) = C_\la \, x_+^{2(\la-2)},
$$
with $x_+^2$ equal to $t^2-|\vec x|^2 = t^2-r^2$ on the forward light
cone and to~0 anywhere else, and
$$
C_\la = \frac{1}{2^{2\la-1}\pi\,\Ga(\la)\Ga(\la-1)},
$$
fulfils
$$
G^\la_\ret * G^\mu_\ret = G^{\la+\mu}_\ret;   \quad
\square G^\la_\ret      = G^{\la-1}_\ret;
\quad  G^0_\ret = \dl.
$$

In particular, $x_+^{2(\la-2)}$ is at once seen to have (generally
double) poles with residues concentrated at the origin as the only
singularities. The $G^\la_\ret$ constitute a set of convolution
inverse powers of the d'Alembertian, verifying
$\square^\la G^\la_\ret = \dl(x)$.

\smallskip

In quantum field theory, as stressed in~\cite{ThreeFromBA}, the
relevant set of inverse powers is related to the Feynman propagator
$$
D_F(x) = \frac{-i}{4\pi^2(t^2 - r^2 - i\eps)}.
$$
We therefore focus on $(t^2-r^2\pm i\eps)^{\a-2}$, with poles at
$\a = 0,-1,\dots$, with the aim of studying the renormalization by
analytic regularization in the variable $\a$ of the functionals $(t^2-r^2\pm
i\eps)^{-2},(t^2-r^2\pm i\eps)^{-3}$ and so on, ill-defined as
distributions. For a start, we need to compute the residues at the poles.

It is instructive and convenient for the purpose to look in the spirit
of analytic regularization at the
Euclidean $\R^4$ first, in a somewhat unconventional manner. Consider
$\rho^2 := |x|^2 = t^2+r^2$. The singularities of $\rho^{\a}$ are well
known (see~I): simple poles in the regularizing variable $\a$ at
$-4-2k$, for $k = 0,1,\dots$, with residues
\begin{equation}
\Res_{\a=-4-2k} \rho^\a =
\frac{\Om_4\,\Dl^k\dl}{2^k k!\,4.6.8\dots(2+2k)},
\label{eq:poles-go}
\end{equation}
the denominator in the case $k = 0$ being~1; here $\Omega_4=2\pi^2$ is
the area of the sphere in dimension~$4$, and $\Dl$ the Laplacian in
dimension 4. It is possible, and usually done, to define a holomorphic
family of distributions that encodes the pole structure of $\rho^{\a}$
in the same way that $\Phi^\la$ encodes that of $x_+^{\la}$. However,
we consider instead the following \textit{meromorphic} family:
$$
G^\a_\eucl(x) = C_\a\,\rho^{2(\a-2)},
$$
with
$$
C_\a = \frac{e^{-i\pi\a}\Ga(2-\a)}{4^{\a}\pi^2\Ga(\a)}.
$$
Notice that $G^1_\eucl$ is the Green function for the Laplace equation
on $\R^4$; that $\Delta G^\a_\eucl = G^{\a-1}_\eucl$, which is quickly
seen from $\Delta\rho^\mu=\mu(\mu+2)\rho^{\mu-2}$; and that
$G^{-m}_{\eucl}(x) = \Delta^m\dl(x)$; the latter of course is just
another way of writing~\eqref{eq:poles-go}.

It is also true that $G^\a_\eucl * G^\b_\eucl = G^{\a+\b}_\eucl$. For
that, define the Fourier transforms on test functions by
$$
F[\phi](p) \equiv \hat\phi(p) :=
\int\frac{d^4x}{(2\pi)^{2}}\,e^{-ipx}\phi(x),  \quad
F^{-1}[\phi](p) \equiv \check\phi(p) :=
\int\frac{d^4x}{(2\pi)^{2}}\,e^{ipx}\phi(x),
$$
and on distributions by transposition. Then
$Ff\,Fg = (2\pi)^{-2}F(f*g)$ for convolvable distributions. It turns
out \cite[Thm.~5.9]{LiebL} that
\begin{equation}
\hat G^\a_\eucl(p) = (2\pi)^{-2}e^{-i\pi\a}|p|^{-2\a},
\label{eq:Fourier-one}
\end{equation}
a most interesting duality. From this the convolution identity follows.

Now we obtain the poles of $(x^2\pm i\eps)^{\a}$ from the poles of
$\rho^{2\a}$. We follow Gelfand and Shilov
\cite[Ch. III, Secs 2.3, 2.4]{ClassicRussian} in this.
Consider the quadratic forms $g_{\pm}(x) := \pm i\rho^2$. Then
$g_{\pm}^\a = e^{\pm i\pi\a/2} \rho^{2\a}$. Rewrite
equation~\eqref{eq:poles-go} as
$$
\Res_{\a=-2-l} g^\a_\pm =
\frac{-\pi^2\,\Dl^l_{g_{\pm}}\dl}
       {4^l l!(l+1)!\sqrt{(\mp i)^4\det g_{\pm}}},
$$
with $\Dl_{g_{\pm}}$ the Laplacian canonically associated to
$g_{\pm}$, which is $\mp i\Dl$ on this occasion. To be precise, if
${\tilde g}^{ij}$ is the inverse matrix of the quadratic form $g$,
then
$$
\Dl_g := \sum_{i,j} \tilde g^{ij} \,\del_i \,\del_j.
$$
For the forms $g(x) = t^2 - r^2 \pm i\eps(t^2+r^2)$, we find then by
analytic continuation:
$$
\Res_{\a=-2-l} (x^2\pm i\eps)^\a =
\frac{\pm i\pi^2\,\square^l\dl}{4^l l!(l+1)!}.
$$
(This analytic continuation is not to be confused with the one
involved in the definition of the $G^\a$ for $\a$ complex.)

The information on the singularity structure of $(x^2\pm i\eps)^{\a}$
---and of its Fourier transform--- can now be codified in
\textit{causal Riesz distributions} $G^\a_{\pm}$. To wit, we define
\begin{equation}
G^\a_{\pm}(x) :=
\frac{\mp ie^{\mp i\pi\a}\Ga(2-\a)}{4^\a\pi^2\Ga(\a)}
(t^2-r^2\pm i\eps)^{\a-2},
\label{eq:causalRiesz}
\end{equation}
and, sure enough,
$$
G^1_{-}(x) = D_F(x); \qquad G^{-l}_{\pm}(x) = \square^l\dl(x)
$$
for $l\geq0$. Also, $\square G^\a_{\pm}(x) = G^{\a-1}_{\pm}(x)$, just
as for the ordinary Riesz distributions. This is clear from
$$
\square(t^2-r^2\pm i\eps)^{\a-2} =
4(\a-1)(\a-2)(t^2-r^2\pm i\eps)^{\a-3}
$$
valid for $1<\Re\a<2$, and then analytically extended. It follows that
$\square^l f = G^{-l}_{\pm} * f$ for appropriately convolvable $f$; and
$\square^\a$ for complex $\a$ can be defined by
$\square^\a f = G^{-\a}_{\pm}*f$.

We can perform the (covariant, if one wishes) Fourier transforms by
the same method of analytic prolongation from the Fourier transforms
of the $r^{2\a}$. The result is
\begin{equation}
\hat G^\a_\pm(p) = (2\pi)^{-2}e^{\mp i\pi\a}(p^2\mp i\eps)^{-\a},
\label{eq:Fourier-two}
\end{equation}
where $p^2 = E^2 - |\vec p|^2$. For instance,
$\hat G^0_-(p) = 1/4\pi^2$,
$\hat G^1_-(p) = \hat D_F(p) = \frac{-1}{4\pi^2(p^2+i\eps)}$, as
expected~\cite{IZ}. There is still an
interesting duality at work here. Moreover,
$$
G^\a_{\pm} * G^\b_{\pm} = G^{\a+\b}_{\pm}.
$$

In summary, thanks to (rigorous) ``Wick rotation'', the structure of
the causal Riesz distributions $G^\a_{\pm}$ is remarkably simpler than
the structure of the retarded Riesz distributions $G^\a_{\ret}$. It
largely parallels the positive signature case, vindicating Schwinger's
contention on the ``Euclidean'' character of quantum field
theory~\cite{JulianS}.

\smallskip

We turn finally to the renormalization of the functionals
$(t^2-r^2 - i\eps)^{-l}$ with $l\geq2$ from $G^\a_-$. We may define
the extension $[(t^2-r^2 - i\eps)^{-2}]_{\rm AR}$, as a distribution,
to be the second term on the right hand side of the expansion
$$
(t^2-r^2-i\eps)^{\kappa-2} =: \frac{-i\pi^2\dl(x)}{\kappa} +
[(t^2-r^2 - i\eps)^{-2}]_{\rm AR} + O(\kappa).
$$
That is to say,
$$
[(t^2-r^2 - i\eps)^{-2}]_{\rm AR} = \lim_{\kappa\to 0}
\frac{d}{d\kappa} [\kappa(t^2-r^2-i\eps)^{\kappa-2}].
$$
Analogously,
\begin{equation}
(t^2-r^2-i\eps)^{\kappa-3} =: \frac{-i\pi^2\square\dl(x)}{8\kappa} +
[(t^2-r^2 - i\eps)^{-3}]_{\rm AR} + O(\kappa),
\label{eq:high-pole}
\end{equation}
as $\kappa\to 0$ defines $[(t^2-r^2 - i\eps)^{-3}]_{\rm AR}$,
and so on.

% \S 4
\section{Lorentz covariance of the $T$-renormalization}

The action of an element $\La$ of the Lorentz group on $\R^{4n}$ is
given by the tensorial representation
$$
\La^{\ox n} x := (\La x_1,\dots,\La x_n),
$$
to be denoted $\La$ as well, according to custom. The action of the
Lorentz group on functionals is defined by
$$
\<\La f(x), \phi(x)> \equiv \<f(\La x), \phi(x)> :=
\<f(x), \La\phi(x)>,  \quad\mbox{with}\quad
\La\phi(x) := \phi(\La^{-1} x).
$$
It follows that $\<\La f(x),\La^{-1}\phi(x)> = \<f(x),\phi(x)>$.

A Lorentz invariant functional fulfils
\begin{equation}
f(\La x) = f(x).
\label{eq:La-inv}
\end{equation}
(More generally, in the nonscalar case, $f$ would have tensorial and/or
spinorial character and one would have a covariant transformation
$$
f(\La x) = [D(\La)f](x)
$$
with $D(\La)$ a finite dimensional representation of $SL(2,\C)$
---making no notational distinction between belonging to the Lorentz
group and to its cover--- acting on functionals in the obvious way.)

Derivatives will transform according to the (tensor powers of the)
contragredient representation: one has
$$
x^\a\,\del_\a(\La\phi) = x^\a\,\del_\a (\phi\circ \La^{-1}) =
x^\a {[\La^{-1}]^\b}_\a (\del_\b\phi)\circ \La^{-1} =
(\La x)^\b (\del_\b\phi)\circ \La^{-1}.
$$
In particular,
\begin{equation}
x^\a\,\del_\a(\La\phi)(0) = [\La^{-1}x]^\b \,\del_\b\phi(0),
\label{eq:0etLa}
\end{equation}
that is to say $R_0^k\La = \La R_0^k$, and
\begin{equation}
\dl_\a(\La x) = [\La^{-1}x]_\a^\b \,\dl_\b(x).
\label{eq:deletLa}
\end{equation}

Suppose that $f$ is Lorentz invariant and a particular extension
$T_w f$ to the whole of $\R^{4n}$ has been constructed, according to
our scheme. \textit{All} the extensions of $f$ are given by
\begin{equation}
T_w f + \sum_{|\a|\leq k}a^\a\,\dl_{\a},
\label{eq:ambig}
\end{equation}
with $\binom{4n + k}{k}$ coefficients $a^\a$. Our goal is to show
that a Lorentz
invariant extension $T^\cov_wf$ can be obtained within the class of
$T$-extensions, advocated in this series of papers. Namely,
$$
T^\cov_w f(\La x) = T^\cov_w f(x),
$$
with
\begin{equation}
T^\cov_w f = T_w f + \sum_{|\a|=k} a^\a\,\dl_{\a},
\label{eq:forcov}
\end{equation}
for at most $\binom{4n-1 + k}{k}$ coefficients $a^\a$;
so that the ambiguity~\eqref{eq:ambig} in all the smaller orders drops
out.

By a theorem of G{\aa}rding and Lions~\cite{SwFr}, the difference
between two covariant extensions must be of the form $P(\square)\dl$,
where $P(\square)$ is a polynomial in $\square$; in our case, a
monomial.

The proof is by direct computation; since $\La f = f$, we get
\begin{align*}
\<\La(T_w f) - T_w f,\phi>
&= \<T_w f,\La\phi> - \<T_w f, \phi> = \<f,T_w\La\phi> - \<f, T_w \phi>
\\
&= \<f, (1-w) R_0^{k-1}\La\phi + w R_0^k \La\phi >
- \<f, (1-w) R_0^{k-1}\phi + w R_0^k \phi>
\\[2mm]
&\stackrel{\mbox{\tiny\eqref{eq:0etLa}}}{=}
\<f, (1-w)\La R_0^{k-1}\phi + w\La R_0^k\phi > -
\<f, (1-w) R_0^{k-1}\phi + w R_0^k \phi>
\\
&= \<\La f, (1-\La^{-1}w) R_0^{k-1}\phi + \La^{-1}w R_0^k\phi > -
\<f, (1-w) R_0^{k-1}\phi + w R_0^k \phi>
\\[2mm]
&\stackrel{\mbox{\tiny\eqref{eq:La-inv}}}{=}
\<f, (1-\La^{-1}w) R_0^{k-1}\phi + \La^{-1}w R_0^k\phi > - \<f, (1-w)
R_0^{k-1}\phi + w R_0^k \phi>
\\
&= \<f, (w - \La^{-1}w) (R_0^{k-1}\phi - R_0^k \phi)>
\\[1.5mm]
&= \sum_{|\a|=k}\<f, (w - \La^{-1}w)x^\a>
\frac{\del_\a \phi (0)}{\a!}.
\end{align*}
The integral $\<f, (w - \La^{-1}w)x^\a>$ exists under the hypothesis
we have made.

This shows that
$$
\La(T_w f) - T_w f = \sum_{|\a|=k} b^\a(\La) \delta_{\a},
$$
with coefficients
$$
b^\a(\La) = (-)^k \<f, (w - \La^{-1}w)x^\a>,
$$
with $|\a|=k$. One also has
\begin{align*}
\sum_{|\a|=k} b^\a(\La) \delta_{\a}
&= k \sum_{|\a|=k} \del^\a \biggl[ \frac{x^\a}{\a!}
\int_0^1 dt\, \frac{(1-t)^{k-1}}{t^{k+4n}} f\Bigl(\frac{x}{t}\Bigr)
\biggl(w\Bigl(\frac{x}{t}\Bigr) - w\Bigl(\frac{\La x}{t}\Bigr)\biggr)
\biggr]
\\
&\quad + (k+1) \sum_{|\b|=k+1} \del^\b \biggl[ \frac{x^\b}{\b!}
\int_0^1 dt\, \frac{(1-t)^k}{t^{k+4n+1}} f\Bigl(\frac{x}{t}\Bigr)
\biggl(w\Bigl(\frac{\La x}{t}\Bigr) - w\Bigl(\frac{x}{t}
\Bigr)\biggr)\biggr].
\end{align*}

The rest of the proof just follows the steps of the cohomological
argument in~\cite{PoSt}: applying two Lorentz transformations, on use
of~\eqref{eq:deletLa} ---and omitting indices--- one obtains
\begin{equation}
b(\La_1\La_2) = \La_2^{-1}b(\La_1) + b(\La_2),
\label{eq:coc-eq}
\end{equation}
where $\La_2^{-1}$ denotes the tensor antirepresentation. A solution
for this equation is given by
\begin{equation}
b(\La) = (1 - \La^{-1})a
\label{eq:coc-sol}
\end{equation}
with $a\in\R^{4k}$ independent of $\La$. Actually~\eqref{eq:coc-eq} is a group
1-cocycle equation, for $SL(2,\C)$, with values in the space carrying the
contragredient representation, and because of the vanishing of the
the first cohomology group $H^1(SL(2,\C);\R^{4k})$~\cite{Wigner} its
only solutions are of the trivial form~\eqref{eq:coc-sol}.

Now, we conclude that, if $a$ satisfies~\eqref{eq:coc-sol}, then, in
view of~\eqref{eq:deletLa}, formula~\eqref{eq:forcov} gives indeed a Lorentz
invariant renormalization of $f$.

For logarithmic divergences, $b = 0$, and,
as $[\La^{-1}]^{\otimes0} = 1$, one can take $a$ arbitrary (this is
a priori obvious in view of the Lorentz invariance of~$\dl$). The
choice $a = 0$ commends itself.

For higher order divergences, in principle one
solves~\eqref{eq:coc-sol} for $a$ and plugs the (in general non
unique) solution in~\eqref{eq:forcov}. However, following a suggestion
in~\cite{Gudrun}, a wiser course can be devised. For simplicity, we
take $n = 1$ from now on. Then we can assume that $f$ depends only on
$x^2$~\cite{RieGut}. It will be seen that only the symmetric part of
the Lorentz content of
$a^\a$ counts. Consider $\sigma(f) = 6$, i.e. $k=2$, a quadratic divergence.
We revert to a Lorentz quadri-index notation: $|\a|=2,\ \a
\leftrightarrow (\mu_1\mu_2)$. It is
found~\cite{Prangeetal} that the totally symmetric part
$$
a^{(\mu_1\mu_2)} =
- \tquarter \<f, \bigl(x^{\mu_1}x^{\mu_2}x^\rho \del_\rho -
x^2x^{(\mu_1}\del^{\mu_2)} \bigr)w>,
$$
is a possible choice for $a$; this choice is canonical in that
$a^\mu_\mu = 0$. Integrating by parts the previous expression, on
account of $\del_\mu f=2x_\mu f'$, we obtain
$$
a^{(\mu_1\mu_2)} =
\<f, \bigl(x^{\mu_1}x^{\mu_2} - \tquarter x^2g^{\mu_1\mu_2}\bigr)w>.
$$
In~\eqref{eq:forcov} with use of \eqref{eq:imp-EG} we see cancellation
of the first term on the right hand side of this equation, and finally
the canonical expression
$$
\<T^\cov_w f(x), \phi(x)> =
\<f(x), \phi(x) - \phi(0) - w(x)\frac{\square\phi(0)}{8} x^2>,
$$
emerges for the Lorentz-invariant distribution extending a
quadratically divergent Lorentz-invariant functional $f$. This formula
supplants the case $k = 2$ of~\eqref{eq:imp-EG} in practice.

More generally, from the formulae in~\cite{Prangeetal}, we can derive,
for $\sigma(f) = 2m+4$ or $2m+5$:
\begin{equation}
\<T^\cov_w f(x), \phi(x)> =
\biggl<f(x), \phi(x) - \phi(0) - \frac{\square\phi(0)}{8} x^2 -\cdots-
w(x) \xi_m x^{2m}\biggr>,
\label{eq:tapadelperoltris}
\end{equation}
where $$\xi_m:= \frac{2(2m-1)!!}{(2m+2)!!(2m)!}.$$

Concentrate now in $T^\cov_\mu$. In Section~4.2 of~I we proved that
the Euclidean $\<[\rho^{-4-2m}]_{\rm AR}, \phi(x)>$ is given by
$$
\biggl<\rho^{-4-2m}, \phi(x) - \phi(0) - \frac{\Dl\phi(0)}{8}\rho^2
-\cdots -
H(1-\rho)\xi_m\,\Dl^m\phi(0)\rho^{2m}\biggr>.
$$
This expression Wick-rotates into
$\<[(x^2\pm i\eps)^{-2-m}]_{\rm AR}, \phi(x)>$, which therefore is
given by
$$
\biggl<(x^2\pm i\eps)^{-2-m}, \phi(x) - \phi(0) -
\frac{\square\phi(0)}{8}x^2 -\cdots -
H(1-\rho)\xi_m\square^m\phi(0)(x^2)^{m}\biggr>.
$$
The conclusion is that $[(x^2\pm i\eps)^{-2-m}]_{\rm AR} =
T^\cov_{\mu=1}(x^2\pm i\eps)^{-2-m}$. For negative powers of the
Feynman propagator, Bollini, Giambiagi and Gonzales
Dom\'{\i}nguez's analytic regularization
and our canonical covariant renormalization using the
\textit{improved} subtraction $T_{\mu=1}$ give one and the same
result.

This coincidence is extended to $T_\mu$ for all values of $\mu$ by
introduction of a 't Hooft factor in the definition of
$[(x^2\pm i\eps)^{-2-m}]_{\rm AR}$. The procedure will be clear from
the examples in the next section.

% \S 5
\section{Computing examples}

The singularities of the powers of $D_F$ are concentrated at the
origin, so that the improved method of Epstein and Glaser is directly
applicable here. Consider first the $T_\mu$-renormalization of
$(t^2-r^2-i\eps)^{-2}$, corresponding to the ``fish'' diagram in the
$\varphi^4_4$ model. We use the notations
$[f]_{\rm R} := [f]_{{\rm R},\mu}:=T^\cov_{\mu}f$.
From~\eqref{eq:tapadelperol} we obtain, in full analogy with the
Euclidean case (see Sec.~3 of~I):
$$
[(x^2-i\eps)^{-2}]_{{\rm R},\mu} = \thalf\del_\nu
\biggl[x^\nu\frac{\log\mu^2(x^2-i\eps)}{(x^2-i\eps)^2}\biggr].
$$
This is the very same result coming from analytical regularization:
just check
$$
\mu^{2\kappa}(x^2-i\eps)^{\kappa-2} =
\frac{\mu^{2\kappa}}{2\kappa}\del_\nu[x^\nu(x^2-i\eps)^{\kappa-2}],
$$
and expand in $\kappa$ the right hand side. In conclusion:
$$
[(t^2-r^2-i\eps)^{-2}]_{{\rm R},\mu} =
[(t^2-r^2-i\eps)^{-2}]_{{\rm AR},\mu},
$$
or $[(D_F)^2]_{\rm R} = [(D_F)^2]_{\rm AR}$, with this generalized
definition of $[\cdot]_{\rm AR}$.

Consider now the ``sunset'' diagram in the same model. Have another
look at Sec.~3 of~I; one obtains
$$
\bigl[(x^2-i\eps)^{-3}\bigr]_{{\rm R},\mu} =
3 \sum_{|\b|=3} \del_\b \biggl[
\frac{x^\b}{\b!}\frac{\log(\mu^2(x^2-i\eps))}{(x^2-i\eps)^{3}}
\biggr] - \frac{3i\pi^2}{8}\,\square\dl(x).
$$
Analogously, one checks by a longer but straightforward calculation:
$$
\mu^{2\kappa}(x^2-i\eps)^{\kappa-3} =
\frac{3\mu^{2\kappa}}{2\kappa(1-3\kappa+2\kappa^2)}
\,\sum_{|\b|=3} \del_\b
\Bigl(\frac{x^\b(x^2-i\eps)^{\kappa-3}}{\b!}\Bigr).
$$
It is clear from \eqref{eq:high-pole} that
$$
3\,\del_\b \Bigl(\frac{x^\b(x^2-i\eps)^{\kappa-3}}{\b!}\Bigr)
= - \frac{i\pi^2}{4}\,\square\dl(x),
$$
from which
$$
\mu^{2\kappa}(x^2-i\eps)^{\kappa-3} =
\frac{-i\pi^2\square\dl(x)}{8\kappa} +
\bigl[(x^2-i\eps)^{-3}\bigr]_{{\rm R},\mu} + O(\kappa).
$$
Therefore $[(D_F)^3]_{\rm R} = [(D_F)^3]_{\rm AR}$.

For higher order powers of $G_F$ similar arguments show that
$$
[(D_F)^l]_{\rm R} = [(D_F)^l]_{\rm AR}
$$
is true generally for $l \geq 2$.

\smallskip
Powers of the massive propagator $D^m_F$ are renormalizable in our standard
way, by use of~\eqref{eq:bas-Lagr}, now applicable, and automatically
Lorentz-covariance preserving. The scaling degree of the (powers of)
propagators is the same in the massive and in the massless cases. One
routinely finds, for instance,
$$
[(D^m_F)^2]_{\rm R}(x) =
-\frac{m^2}{32\pi^4} \del_\mu x^\mu \biggl(
\frac{K_1^2(m\sqrt{-x^2+i\eps})-K_0(m\sqrt{-x^2+i\eps})
K_2(m\sqrt{-x^2+i\eps})}{-x^2+i\eps} \biggr).
$$

% \S 6
\section{BPHZ renormalization revisited}

It is well known that for zero-mass models, the basic BPHZ scheme runs
into trouble. This is due to the failure of $\del^\mu\hat f(0)$ to
exist for $|\mu|=k$, on account of the infrared problem. Now, one can
try subtraction at some suitable external momentum $q\neq0$, providing
a mass scale. It is patent that this last subtraction will introduce
in the Minkowskian context a noncovariance. This prompted Lowenstein
and Zimmermann to introduce their ``soft mass
insertions''~\cite{LowZim}; but that BPHZL method is quite awkward in
practice.

A far simpler solution to the problem is now available to us. It
comes from the observation in~I that the BPHZ method is ancillary to
Epstein and Glaser's: from the definitions
\begin{equation}
\<F[R^k_0f], F^{-1}[\phi]> = \<F[f], F^{-1}[R^k_0\phi]>.
\label{eq:deep-mystery}
\end{equation}
An expression such as $F[f]$ is \textit{not} a priori meaningless: it
is a well defined functional on the linear subspace of Schwartz
functions $\phi$ whose first moments $\int p^\a\phi(p)\,d^d p$ up to
order $k+1$ happen to vanish: this is the Fourier counterpart of the
space of distributions on configuration space acting on Schwartz test
functions vanishing up to order $k+1$ at the origin.

Now, one has
$$
(x^\mu \phi)\check{\ }(p) = (-i)^{|\mu|}\,\del^\mu\check\phi(p),
$$
where $\mu$ denotes a multiindex; so that, in particular,
$$
(x^\mu)\check{\ }(p) = (-i)^{|\mu|} (2\pi)^{d/2}\,\del^\mu\dl(p).
$$
Also,
$$
\del_\mu\phi(0) = (-i)^{|\mu|} (2\pi)^{-d/2} \<p^\mu, \check\phi>.
$$
{}From this, with an integration by parts in the right hand side
of~\eqref{eq:deep-mystery}, we conclude
$$
\<F[R^k_0f], F^{-1}[\phi]> = \<R^k_0 F[f], F^{-1}[\phi]>;
$$
that is to say, $F$ and $R^k_0$ \textit{commute}. Thus the BPHZ
subtraction rule in momentum space is equivalent to~\eqref{eq:bas-Lagr}.

Now use~\eqref{eq:tapadelperoltris} instead of~\eqref{eq:bas-Lagr},
with employment of $w(x)=\exp(-iqx)$, with $q\neq0$, which, as
discussed earlier, is a perfectly good infrared regulator. From that
follows the simple, obviously covariant, rule:
$$
T^\cov_q f(p) =
f(p) - \frac{\square f(0)p^2}{8} -\cdots- \xi_m\square^m f(q)\,(p^2)^m.
$$
for a Feynman amplitude $f$ in momentum space. Note
$\square^m T^\cov_q f(q) = 0$. The difference between two of these
recipes is, as it should be, a Lorentz-invariant polynomial in $p$, of
degree the divergence index.

% \S 7
\section{Outlook}

The ``missing link'' between the Epstein--Glaser subtraction method
and the literature on prolongation of distributions found in~I has
here been extended to the Minkowskian context. Before rendering in the
language of $T$-renormalization the full complexity of the
construction of time-ordered products, and the main result of
perturbative renormalization theory, one needs to handle the
combinatorial aspects of diagrams with subdivergences. This we do in
the next paper of the series~III, using a variant of the
Connes--Kreimer Hopf algebraic paradigm.

\bigskip

\subsection*{Acknowledgements}

We thank Christian Brouder and Joe V\'arilly for helpful comments.
Support from the Vicerrector\'{\i}a de Investigaci\'on of the
Universidad de Costa Rica is gratefully acknowledged, as well as from
Universit\'e de Provence in the early stage of the work.


\bigskip



\begin{thebibliography}{34}

\bibitem{EG}
H. Epstein and V. Glaser,
Ann. Inst. Henri Poincar\'e {\bf XIXA} (1973) 211.
%%CITATION = AHPAA,A19,211;%%

\bibitem{Carme}
J. M. Gracia-Bond\'{\i}a,
Math. Phys. Anal. Geom. {\bf 6} (2003) 59.

\bibitem{EllRen}
R. Stora,
\textsl{A note on elliptic perturbative renormalization on a compact
manifold}, unpublished notes, CERN \& LAPP-TH.

\bibitem{PoSt}
G. Popineau and R. Stora,
\textsl{A pedagogical remark on the main theorem of perturbative
renormalization theory},
unpublished preprint, CPT, CERN \& LAPP-TH (1982).

\bibitem{Scharf}
G. Scharf,
\textit{Finite Quantum Electrodynamics: the Causal Approach},
Springer, Berlin, 1995.

\bibitem{Prangeetal}
K. Bresser, G. Pinter and D. Prange,
\textsl{Lorentz invariant renormalization in causal perturbation
theory},
hep-th/9903266, DESY, 1999.
%%CITATION = HEP-TH 9903266;%%

\bibitem{PrangeII}
D. Prange,
\textsl{Lorentz covariance in Epstein--Glaser renormalization},
hep-th/9904136, DESY, 1999.
%%CITATION = HEP-TH 9904136;%%

\bibitem{ThreeFromBA}
C. G. Bollini, J. J. Giambiagi and A. Gonzales Dom\'{\i}nguez,
Nuovo Cim. {\bf XXXI} (1964) 550.
%%CITATION = NUCIA,31,550;%%

\bibitem{Riesz}
M. Riesz,
Acta Math. {\bf 81} (1949) 1.
%%CITATION = ACMAA,81,1;%%

\bibitem{Jager}
E. M. de Jager,
in \textit{Mathematics Applied to Physics}, E. Roubine, ed.,
Springer, Berlin, 1970.

\bibitem{Kanwal}
R. P. Kanwal,
\textit{Generalized Functions: Theory and Technique},
Birkh\"auser, Boston, 1998.

\bibitem{Flora}
J. M. Gracia-Bond\'{\i}a,
\textsl{Improved Epstein--Glaser renormalization in coordinate space
III. The Hopf algebra of Feynman graphs}, San Jos\'e, 2003.

\bibitem{DirkOriginal}
D. Kreimer,
Adv. Theor. Math. Phys. {\bf 2} (1998) 303.
%%CITATION = 00203,2,303;%%

\bibitem{CKI}
A. Connes and D. Kreimer,
Commun. Math. Phys. {\bf 210} (2000) 249.
%%CITATION = HEP-TH 9912092;%%

\bibitem{Hektor}
H. Figueroa and J. M. Gracia-Bond\'{\i}a,
{\sl The uses of Connes--Kreimer's algebraic formulation of
renormalization theory}, hep-th/0301015, San Jos\'e, 2003, to appear in Int.
J. Mod. Phys. A.

\bibitem{Schw}
L. Schwartz,
\textit{Th\'eorie des distributions},
Hermann, Paris, 1966.

\bibitem{IZ}
C. Itzykson and J.-B. Zuber,
\textit{Quantum Field Theory},
McGraw-Hill, New York, 1980.

\bibitem{OS}
O. Steinmann,
\textit{Perturbation Expansions in Axiomatic Field Theory},
Lecture Notes in Physics 11, Springer, Berlin, 1971.

\bibitem{Ricardo}
R. Estrada,
Int. J. Math. \& Math. Sci. {\bf 21} (1998) 625.
%%CITATION = 00145,21,625;%%

\bibitem{Odysseus}
R. Estrada, J. M.
Gracia-Bond\'{\i}a and J. C. V\'arilly,
Commun. Math. Phys. {\bf 191} (1998) 219.
%%CITATION = FUNCT-AN 9702001;%%

\bibitem{PrangeA}
D. Prange,
J. Phys. A {\bf 32} (1999) 2225.
%%CITATION = HEP-TH 9710225;%%

\bibitem{Hungarian}
R. Estrada,
Publ. Math. Debrecen {\bf 61} (2002) 1.
%%CITATION = PUMAA,61,1;%%

\bibitem{OSphi}
O. Steinmann,
Commun. Math. Phys. {\bf 152} (1993) 627.
%%CITATION = CMPHA,152,627;%%

\bibitem{Hoermander}
L. H\"ormander,
\textit{The Analysis of Partial Differential Operators I},
Springer, Berlin, 1983.

\bibitem{Gudrun}
G. Pinter, ``Epstein--Glaser renormalization: finite renormalizations,
the $\Sf$-matrix of $\Phi^4$ theory and the action principle'',
Doktorarbeit, DESY, 2000.

\bibitem{LiebL}
E. H. Lieb and M. Loss,
\textit{Analysis},
Graduate Studies in Mathematics {\bf 14},
Amer. Math. Soc., Providence, RI, 1997.

\bibitem{ClassicRussian}
I. M. Guelfand and G. E. Chilov,
\textit{Les distributions I},
Dunod, Paris, 1972.

\bibitem{JulianS}
J. S. Schwinger,
Proc. Natl. Acad. Sci. USA {\bf 44} (1958) 956.
%%CITATION = PNASA,44,956;%%

\bibitem{SwFr}
L. G{\aa}rding and J.-L. Lions,
Nuovo Cim. Suppl. {\bf XXIV} (1959) 9.

\bibitem{Wigner}
E. Wigner,
Ann. Math. {\bf 40} (1939) 149.
%%CITATION = ANMAA,40,149;%%

\bibitem{RieGut}
A. Rieckers and W. G\"uttinger,
Commun. Math. Phys. {\bf 7} (1968) 190.

\bibitem{LowZim}
J. H. Lowenstein and W. Zimmermann,
Nucl. Phys. {\bf B86} (1975) 77.
%%CITATION = NUPHA,B86,77;%%


\end{thebibliography}
\end{document}